\documentclass[conference]{IEEEtran}
\IEEEoverridecommandlockouts
\usepackage{cite}
\usepackage{amsmath,amssymb,amsfonts}
\usepackage{algorithmic}
\usepackage{algorithm}
\usepackage{graphicx}
\usepackage{textcomp}
\usepackage{xcolor}
\usepackage{hyperref}
\usepackage{booktabs}
\usepackage{multirow}
\usepackage[table]{xcolor} 
\usepackage{colortbl}
\usepackage{subfig, siunitx, adjustbox}
\def\BibTeX{{\rm B\kern-.05em{\sc i\kern-.025em b}\kern-.08em
    T\kern-.1667em\lower.7ex\hbox{E}\kern-.125emX}}

\begin{document}

\title{Blockage-Aware Multi-RIS Sensing and Optimization for mmWave Smart Radio Environments
}

% \author{Anonymous Authors}
\author{\IEEEauthorblockN{Sehyun Ryu\textsuperscript{1,3} and Hyun Jong Yang\textsuperscript{2,3}}
\IEEEauthorblockA{\textsuperscript{1}\textit{Department of Electrical Engineering, POSTECH, Republic of Korea}\\
\textsuperscript{2}\textit{Department of Electrical and Computer Engineering, Seoul National University, Republic of Korea}\\
\textsuperscript{3}\textit{Institute of New Media and Communications, Seoul National University, Republic of Korea} \\
 sh.ryu@postech.ac.kr, hjyang@snu.ac.kr}

\thanks{Accepted to IEEE GLOBECOM 2026.}
\thanks{This work was supported by grants from the Institute of Information \& Communications Technology Planning \& Evaluation (IITP), funded by the Korean government (MSIT): Grant No. RS-2026-25523626, “Development of End-to-End Service Performance Guarantee Framework Technology for 6G,” and Grant No. RS-2024-00428780, “6G Cloud Research and Education Open Hub.” (Corresponding author: Hyun Jong Yang.)}
}

\maketitle

\begin{abstract}

% Multiple reconfigurable intelligent surfaces (RISs) can provide spatially diverse bypass paths to mitigate mmWave blockage, but exploiting this diversity requires knowledge of which RIS-assisted paths are available to each user equipment (UE).
% The resulting challenge is to identify user-specific RIS availability with low sensing overhead before jointly configuring the base-station (BS) precoder and RIS phases.
% To address this challenge, we propose a sensing-driven blockage-aware multi-RIS optimization framework that first detects user-specific available RIS sets and then jointly optimizes the BS precoder and RIS phases for weighted sum-rate (WSR) maximization.
% The BS transmits short per-RIS-indexed synchronization sequences with a cyclic prefix, enabling each user equipment to resolve the availability of individual RIS-assisted paths via low-complexity energy detection under multipath and inter-RIS interference.
% Based on the detected RIS availability sets, we develop a Closed-form Riemannian Phase Alignment (CRPA) algorithm for multi-RIS WSR maximization.
% CRPA yields unit-modulus-preserving closed-form phase updates through a block-MM surrogate; it guarantees
% monotonic improvement for continuous RIS phases and employs an objective-safeguarded quantized update for practical discrete phase shifts.
% Simulations demonstrate accurate per-RIS path availability sensing and improved WSR and convergence performance over existing multi-RIS baselines under mmWave-realistic blockage conditions.

Multiple reconfigurable intelligent surfaces (RISs) can provide spatially diverse bypass paths against mmWave blockage, but exploiting this diversity requires low-overhead identification of the RIS-assisted paths available to each user equipment (UE) before jointly configuring the base-station (BS) precoder and RIS phases.
A sensing-driven blockage-aware multi-RIS optimization framework is proposed to first detect user-specific available RIS sets and then maximize the weighted sum-rate (WSR).
The BS transmits short per-RIS-indexed synchronization sequences with a cyclic prefix, enabling low-complexity energy detection under Rician fading and inter-RIS interference.
Based on the detected sets, a Closed-form Riemannian Phase Alignment (CRPA) algorithm is developed, yielding unit-modulus-preserving closed-form updates through a block-MM surrogate, monotonic improvement for continuous phases, and objective-safeguarded updates for quantized phases.
Simulations demonstrate accurate per-RIS availability sensing and improved WSR and convergence over existing multi-RIS baselines under mmWave-realistic blockage conditions.
\end{abstract}


\section{Introduction}
Millimeter-wave (mmWave) communications have been widely recognized as a key technology for high-capacity wireless systems \cite{RAPPAPORT2019}.
However, due to weak diffraction and strong dependence on line-of-sight propagation, mmWave links are highly susceptible to blockage.
Reconfigurable intelligent surfaces (RISs) have emerged as a promising means of mitigating this limitation by
programmably shaping incident electromagnetic waves and reconfiguring the wireless propagation environment \cite{WU2019,CHEN2020}.
In this sense, RISs are a key enabler of smart radio environments, where the propagation medium is no longer treated as a fixed and uncontrollable entity but as a configurable resource for improving coverage, spectral
efficiency, and reliability.

A large body of work has studied the joint optimization of the active BS precoder and passive RIS phase shifts. 
Early RIS designs focused on transmit-power minimization \cite{WU2019}, QoS-aware beamforming \cite{LIU2021}, and received-power maximization through joint optimization of beamforming, RIS phases, and deployment geometry \cite{CHENG2022}.
Weighted sum-rate (WSR) maximization has also become a central objective in multi-user RIS-assisted systems, including multi-user multiple-input single-output (MU-MISO) formulations \cite{GUO2020} and mmWave scenarios based on fractional programming and alternating optimization \cite{DAMPAHALAGE2021}.
These studies have subsequently been extended from single-RIS settings to multi-RIS architectures.
For example, double-RIS channel estimation and passive beamforming were considered in \cite{YOU2021}, WSR maximization for cooperative multi-RIS transmission was investigated in \cite{LI2020}, Riemannian manifold optimization was applied to double-RIS multi-user multiple-input multiple-output (MU-MIMO) mmWave systems in \cite{NIU2022}, and auxiliary-variable-based beamforming was developed for multiple RIS-assisted communication systems in \cite{MA2022}.

% Since RIS was originally proposed as a means to overcome blockage, a number of studies have also focused on cases where the RIS paths themselves are blocked.
% Early work \cite{KUMAR2021} investigated beamformer optimization under random link blockages across different blockage combinations.
% Subsequent studies primarily relied on statistical assumptions about blockage.
% For example, \cite{JIAO2022} employed stochastic learning to capture crucial blockage patterns and designed robust beamforming against uncertain path blockages, while \cite{ZHOU2021} proposed a learning-based robust beamforming scheme for RIS-aided mmWave systems under random blockages.
% More recently, \cite{SARKER2024} optimized the precoder and phase shifts to maximize WSR in mmWave massive MIMO with mobile RIS under known blockage conditions, but without addressing how to detect blockage in practice.
% In contrast, \cite{YANG2024} proposed a binary Neyman–Pearson (NP) detection method to sense and optimize blockage in a single-RIS system.
% However, in multi-RIS scenarios, the superposition of multipath components from different RIS panels makes per-RIS signal isolation difficult, rendering NP-style binary detection infeasible.

Multi-RIS deployments are particularly attractive for mmWave smart radio environments because they provide multiple spatially diverse programmable propagation paths.
Compared with a single-RIS deployment, a multi-RIS system can offer alternative reflected links for different users and thus more flexible environment configuration under blockage.
This flexibility, however, can be exploited only if the network is aware of which RIS-assisted paths are currently available.
In practice, an RIS-assisted path may itself be blocked by moving obstacles, human bodies, vehicles, or environmental changes. If all RIS panels are assumed to be available, the resulting precoder and RIS configuration may allocate resources to blocked paths and suffer substantial performance degradation.

Blockage-aware RIS communications have therefore attracted increasing attention.
Early work studied beamformer optimization under random link blockages across different blockage combinations \cite{KUMAR2021}.
Subsequent studies considered stochastic or learning-based robust beamforming under uncertain blockage models \cite{JIAO2022,ZHOU2021}.
More recently, joint precoder and RIS phase optimization was studied for mmWave massive MIMO systems with mobile RISs under known blockage conditions \cite{SARKER2024}, and a binary Neyman--Pearson (NP) detection approach was proposed for sensing and optimizing blockage in a single-RIS mobile mmWave MIMO system \cite{YANG2024}.
These works have significantly advanced blockage-aware RIS beamforming, but they do not directly address a key requirement of multi-RIS smart radio environments: each UE must resolve the availability of multiple simultaneously deployed RIS-assisted paths from downlink observations before the network can configure the RIS phases and BS precoder.
In multi-RIS systems, this is nontrivial because the received pilot contains superposed contributions from multiple RIS panels, cross-illumination, and inter-index leakage.

This paper proposes a sensing-driven blockage-aware multi-RIS optimization framework for mmWave smart radio environments.
The BS transmits short per-RIS-indexed synchronization sequences with a cyclic prefix, and each UE performs low-complexity energy detection on the corresponding matched-filter outputs to identify its user-specific available RIS set.
Unlike robust beamforming methods that assume statistical blockage information or known blockage states, the proposed framework explicitly senses the availability of individual RIS-assisted paths and uses the detected sets for subsequent multi-RIS configuration.
Based on these detected sets, the BS precoder and RIS phase shifts are jointly optimized for WSR maximization using a Closed-form Riemannian Phase Alignment (CRPA) algorithm.
CRPA derives unit-modulus-preserving phase updates from a block-MM surrogate, guarantees monotonic improvement for continuous RIS phases, and employs an objective-safeguarded quantized update for practical discrete phase shifts.

The main contributions are summarized as follows:
\begin{itemize}
\item A per-RIS-indexed downlink sensing protocol is proposed
to identify user-specific available RIS sets under Rician fading
and inter-RIS interference.
\item Blockage-aware multi-RIS WSR maximization is formulated using the detected RIS sets.
\item CRPA is developed by reducing the quadratic MM subproblem to closed-form unit-modulus phase alignment, with monotonic improvement for continuous phases and safeguarded updates for quantized phases.
\item Simulations validate accurate RIS-path sensing and improved WSR and convergence over multi-RIS baselines.
\end{itemize}

% Multi-RIS를 활용한 MU-MIMO 시스템. obstacles에 의해 blockage가 발생하여 UE마다 사용 가능한 RIS panel set이 달라진다. 

\section{System Model and Problem Formulation}
\begin{figure}[t]
    \centering
    \includegraphics[width=0.9\linewidth]{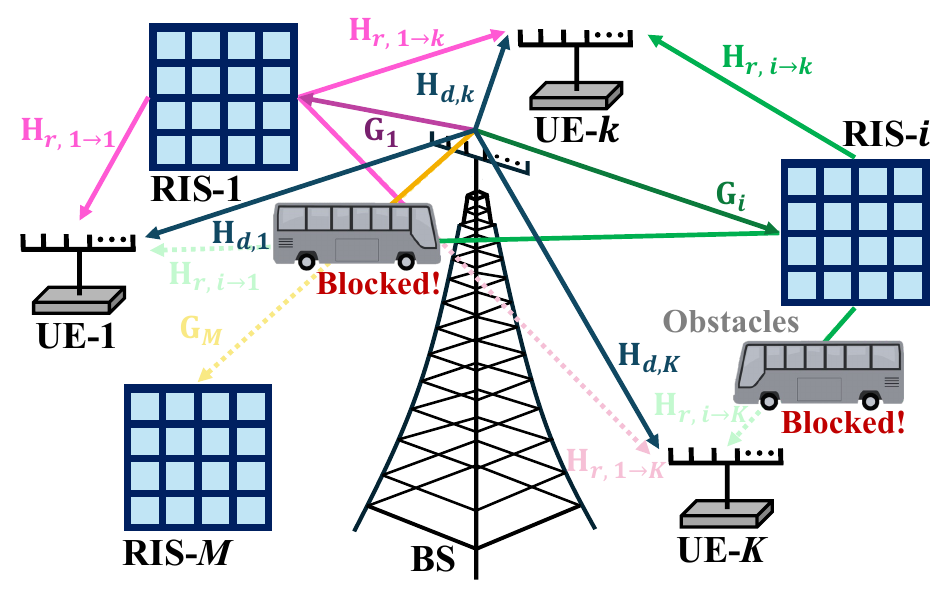}
    \vspace{-7pt}
    \caption{Multi-RIS aided mmWave MU--MIMO downlink system.}
    \vspace{-8pt}
    \label{fig:Fig1}
\end{figure}

This paper considers a downlink mmWave MU--MIMO multi-RIS system with $K$ users and $M$ RIS panels, where RIS-$i$ consists of $N_i$ reflecting elements (Fig.~\ref{fig:Fig1}).
Each RIS operates as a passive phase-only surface with constant-modulus reflection coefficients.
The BS employs $N_t$ antennas, while UE-$k$ employs $N_{r,k}$ antennas.

The system employs orthogonal frequency-division multiplexing (OFDM) downlink and adopts an equivalent per-subcarrier flat-fading model after cyclic-prefix (CP) removal.
The BS appends a CP to each pilot symbol with a length exceeding the
maximum delay spread, thereby preserving cyclic correlation under
frequency-selective propagation, as detailed in Section~\ref{Sec:detect}.

Let $\mathcal A_k \subseteq \{1,\dots,M\}$ denote the \emph{true available RIS set} for UE-$k$, i.e., the set of RIS panels whose cascaded BS$\!\to$RIS-$i$$\!\to$UE-$k$ paths are unblocked.
From the downlink synchronization signals, UE-$k$ obtains a \emph{detected available RIS set} $\widehat{\mathcal A}_k$.
Throughout the paper, $\mathcal A_k$ denotes the unknown ground-truth set, whereas $\widehat{\mathcal A}_k$ denotes its detected counterpart used for optimization.

\subsection{Per-RIS Availability Detection}
Each RIS panel is assigned a distinct sequence index, as detailed in Section~\ref{Sec:detect}.
Let $Z_{k,i}$ denote the matched-filter output at UE-$k$ for RIS index $i$.
The detected available RIS set is obtained as
\begin{equation}
\widehat{\mathcal A}_k =
\{ i\in\{1,\ldots,M\}: |Z_{k,i}|^2 \ge \widehat{\tau}_{k,i} \},
\end{equation}
where $\widehat{\tau}_{k,i}$ is the estimated UE--RIS-specific detection threshold.
Detection accuracy is measured by
\begin{equation}
J_k =
\frac{|\widehat{\mathcal A}_k\cap\mathcal A_k|}
{|\widehat{\mathcal A}_k\cup\mathcal A_k|},
\end{equation}
with $J_k=1$ when both sets are empty.

\subsection{WSR Maximization}
For RIS-$i$, let $\mathbf u_i \in \mathbb C^{N_i}$ denote its reflection coefficient vector with
$|[\mathbf u_i]_m|=1,\ \forall m$.
For UE-$k$, the true physical effective channel is determined by the true available RIS set $\mathcal A_k$ as
\begin{equation}
\label{eq:Heff-true}
\mathbf H_{k,\mathrm{eff}}
=
\mathbf H_{d,k}
+
\sum_{i\in\mathcal A_k}
\mathbf H_{r,i\to k}\,\mathrm{diag}(\mathbf u_i)\,\mathbf G_i ,
\end{equation}
where $\mathbf H_{d,k}$, $\mathbf G_i$, and $\mathbf H_{r,i\to k}$ denote the direct BS--UE, BS--RIS-$i$, and RIS-$i$--UE-$k$ channels, respectively.

Since $\mathcal A_k$ is unknown to the network during operation, the optimizer constructs an estimated effective channel from the detected available RIS set $\widehat{\mathcal A}_k$:
\begin{equation}
\widehat{\mathbf H}_{k,\mathrm{eff}}
=
\mathbf H_{d,k}
+
\sum_{i\in\widehat{\mathcal A}_k}
\mathbf H_{r,i\to k}\,\mathrm{diag}(\mathbf u_i)\,\mathbf G_i .
\end{equation}
The optimizer uses $\widehat{\mathbf H}_{k,\mathrm{eff}}$, while performance is evaluated using the true channel $\mathbf H_{k,\mathrm{eff}}$.

With a linear precoder $\mathbf F=[\mathbf f_1,\dots,\mathbf f_K] \in \mathbb C^{N_t\times K}$ satisfying $\|\mathbf F\|_F^2 \le P$, the actual received signal at UE-$k$ is
\begin{equation}
\mathbf y_k
=
\mathbf H_{k,\mathrm{eff}} \mathbf F \mathbf s + \mathbf n_k,
\end{equation}
where $\mathbf s\in\mathbb C^K$ and $\mathbf n_k \sim \mathcal{CN}(0,\sigma^2 \mathbf I_{N_{r,k}})$.

Since the true available set $\mathcal A_k$ is not known during optimization, the BS and RIS phases are designed using the detected effective channels $\{\widehat{\mathbf H}_{k,\mathrm{eff}}\}_{k=1}^K$.
The resulting WSR maximization problem is
\begin{align}
\label{eq:wsr}
& \max_{\mathbf F,\{\mathbf u_i\}}
\sum_{k=1}^K \omega_k
\log\det\!\big(\mathbf I_{N_{r,k}} +
\widehat{\mathbf R}_k^{-1}
\widehat{\mathbf H}_{k,\mathrm{eff}} \mathbf f_k \mathbf f_k^{\!H}
\widehat{\mathbf H}_{k,\mathrm{eff}}^{\!H}\big) \\
& \text{s.t.}\;\;
\|\mathbf F\|_F^2\le P,\quad
|[\mathbf u_i]_m|=1,\ \forall i,m, \nonumber
\end{align}
where
\begin{equation}
\widehat{\mathbf R}_k
=
\sigma^2\mathbf I_{N_{r,k}} +
\sum_{j\neq k}
\widehat{\mathbf H}_{k,\mathrm{eff}}
\mathbf f_j \mathbf f_j^{\!H}
\widehat{\mathbf H}_{k,\mathrm{eff}}^{\!H}.
\end{equation}
After optimization, the achieved WSR is evaluated over the true physical channels $\{\mathbf H_{k,\mathrm{eff}}\}_{k=1}^K$.

\section{Per-RIS Indexed Synchronization and Blockage Detection}
\label{Sec:detect}

\begin{figure}[t]
    \centering
    \includegraphics[width=0.86\linewidth]{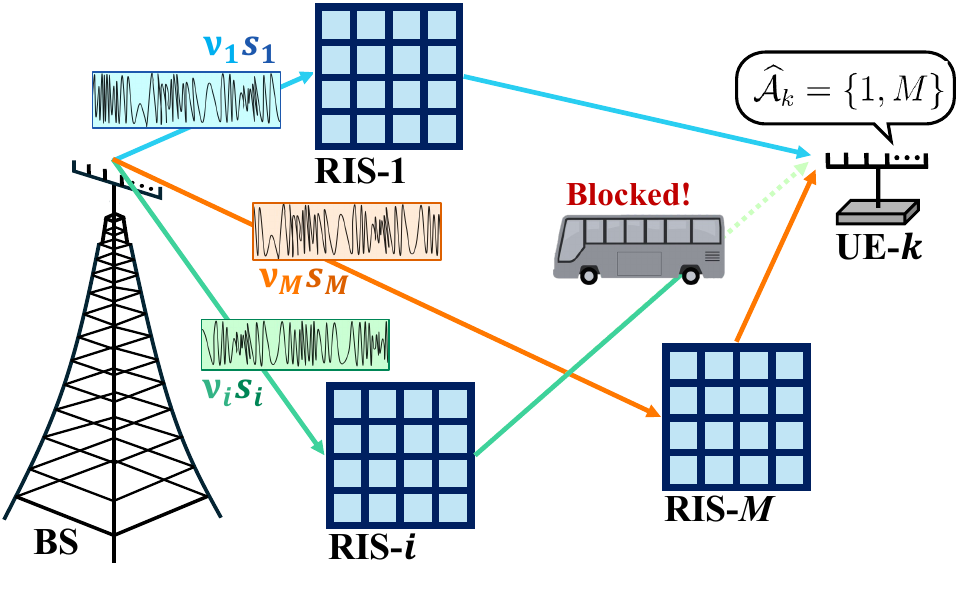}
    \vspace{-7pt}
    \caption{Per-RIS indexed synchronization for constructing a user-specific detected available RIS set.}
    \vspace{-7pt}
    \label{fig:Fig2}
\end{figure}

Fig.~\ref{fig:Fig2} illustrates the per-RIS indexing, where each RIS panel is assigned a distinct cyclic shift of a root Zadoff--Chu (ZC) sequence of length $\ell$:
\begin{align}
s[n]
&= \exp\!\Big(-j \tfrac{\pi q n(n+1)}{\ell}\Big),\\
s_i[n]
&\triangleq s[(n+i)\!\!\mod \ell],\quad i=1,\dots,M.
\end{align}
If $M\le\ell$, all RIS panels are indexed within one pilot symbol; otherwise, $T_p=\lceil M/\ell\rceil$ pilot symbols or multiple ZC roots are used.
Each OFDM pilot includes a CP to support cyclic correlation after CP removal.

During sensing, the BS transmits a superposition of RIS-indexed pilot components
\begin{equation}
\label{eq:tx-superposition}
\mathbf x[n]=\sum_{i=1}^{M}\mathbf v_i s_i[n],\qquad
\sum_{i=1}^{M}\|\mathbf v_i\|_2^2\le P_{\mathrm{pilot}},
\end{equation}
where $\mathbf v_i$ is the sensing beam associated with RIS-$i$;
equal pilot power is used unless otherwise stated.
The sensing beams provide directional isolation such that
$\|\mathbf G_j\mathbf v_i\|_2^2
\ll
\|\mathbf G_i\mathbf v_i\|_2^2$ for $j\neq i$.
Direct BS--UE leakage of RIS-indexed pilots is assumed sufficiently weak under directional isolation and is absorbed into the aggregate disturbance.

The index $i$ does not imply that RIS-$i$ generates a pilot; rather, the BS transmits a pilot component associated with RIS-$i$, while RIS-$i$ reflects it using the known sensing state.
RIS-$i$ applies a predefined sensing reflection vector
$\mathbf u_i^{(0)}\in\mathbb C^{N_i}$ satisfying
$|[\mathbf u_i^{(0)}]_m|=1$, which is fixed and known to the BS and UEs.
A simple choice is $\mathbf u_i^{(0)}=\mathbf 1_{N_i}$.

Given $\mathbf u_i^{(0)}$, define the RIS-$i$ sensing channel observed by UE-$k$ as
\begin{equation}
\mathbf H^{\mathrm{sen}}_{k,i}
=
\begin{cases}
\mathbf H_{r,i\to k}\mathrm{diag}(\mathbf u_i^{(0)})\mathbf G_i,
& i\in\mathcal A_k,\\
\mathbf 0, & i\notin\mathcal A_k.
\end{cases}
\end{equation}
Under the directional spatial-isolation and per-subcarrier flat-fading
approximations, the post-CP signal used for RIS sensing is
\begin{equation}
\label{eq:rx}
\mathbf y_k[n]
=
\sum_{i=1}^{M}
\mathbf H^{\mathrm{sen}}_{k,i}\mathbf v_i s_i[n]
+\boldsymbol{\varepsilon}_k[n]
+\mathbf n_k[n],
\end{equation}
where $\boldsymbol{\varepsilon}_k[n]$ aggregates residual direct leakage,
cross-illumination, and delay-induced interference, and
$\mathbf n_k[n]\sim\mathcal{CN}(\mathbf0,\sigma^2\mathbf I)$.

Since $\mathbf H^{\mathrm{sen}}_{k,i}$ is unknown during sensing, UE-$k$ uses a fixed combiner
$\widetilde y_k[n]=\mathbf w_k^H\mathbf y_k[n]$
with $\|\mathbf w_k\|_2=1$.
The matched-filter statistic for RIS index $i$ is
\begin{equation}
\label{eq:mf-def}
Z_{k,i}
\triangleq
\sum_{n=0}^{\ell-1}s_i^*[n]\widetilde y_k[n],
\end{equation}
which can be decomposed as
\begin{equation}
\label{eq:mf-output}
Z_{k,i}
=
\ell\,\mathbf w_k^H
\mathbf H^{\mathrm{sen}}_{k,i}\mathbf v_i
+
\underbrace{
\sum_{j\ne i}
\rho_{i,j}\mathbf w_k^H
\mathbf H^{\mathrm{sen}}_{k,j}\mathbf v_j
+\varepsilon_{k,i}
}_{I_{k,i}}
+\widetilde n_{k,i},
\end{equation}
where
$\widetilde n_{k,i}\sim\mathcal{CN}(0,\sigma^2\ell)$ and
$\rho_{i,j}\triangleq
\sum_{n=0}^{\ell-1}s_i^*[n]s_j[n]$.
Here, $\varepsilon_{k,i}$ denotes the matched-filtered residual corresponding to $\boldsymbol{\varepsilon}_k[n]$ and is included in the effective disturbance variance.
The desired first term exists only when $i\in\mathcal A_k$.

For each UE--RIS pair $(k,i)$, the hypotheses are
\[
\mathcal H_{1,k,i}:i\in\mathcal A_k
\quad\text{versus}\quad
\mathcal H_{0,k,i}:i\notin\mathcal A_k.
\]
Under $\mathcal H_{0,k,i}$, the desired RIS-$i$ component is absent, while thermal noise and residual inter-index leakage may remain.
Accordingly, the aggregate disturbance is approximated as
\begin{equation}
Z_{k,i}\mid\mathcal H_{0,k,i}
\approx
\mathcal{CN}(0,\sigma_{\mathrm{eff},k,i}^2\ell),
\end{equation}
where $\sigma_{\mathrm{eff},k,i}^2$ includes thermal noise and residual leakage.
Since
$|Z_{k,i}|^2/(\sigma_{\mathrm{eff},k,i}^2\ell)
\sim\operatorname{Exp}(1)$
under $\mathcal H_{0,k,i}$, imposing false-alarm probability $\alpha_{k,i}$ yields the NP threshold
\begin{equation}
\label{eq:np-threshold}
\tau_{k,i}
=
-\sigma_{\mathrm{eff},k,i}^2\ell\ln\alpha_{k,i}.
\end{equation}

In practice, UE-$k$ obtains $T_{\mathrm g}$ disturbance-only samples from existing guard or idle resources in which the $i$-th indexed pilot is absent, adding no pilot overhead.
Correlation with $s_i[n]$ yields
$\{Z_{k,i}^{(\mathrm g)}(t)\}_{t=1}^{T_{\mathrm g}}$ and
\begin{equation}
\label{eq:sigmahat}
\widehat{\sigma}_{\mathrm{eff},k,i}^2
=
\frac{1}{T_{\mathrm g}\ell}
\sum_{t=1}^{T_{\mathrm g}}
\big|Z_{k,i}^{(\mathrm g)}(t)\big|^2.
\end{equation}
The practical threshold is
$\widehat{\tau}_{k,i}
=-\widehat{\sigma}_{\mathrm{eff},k,i}^2\ell\ln\alpha_{k,i}$.
Finally, UE-$k$ constructs its detected available RIS set as
\begin{equation}
\label{eq:Ahat-final}
\widehat{\mathcal A}_k
=
\Big\{
i\in\{1,\dots,M\}:|Z_{k,i}|^2\ge\widehat{\tau}_{k,i}
\Big\}.
\end{equation}

Sensing requires $T_p=\lceil M/\ell\rceil$ OFDM pilot symbols; hence, when $M\le\ell$, all RISs are indexed within one symbol instead of sequential $M$-slot probing.
Per-UE combining and FFT-based correlation cost
$O(N_{r,k}\ell)$ and $O(\ell\log\ell)$, respectively.

\section{Closed-form Riemannian Phase Alignment (CRPA) for Multi-RIS WSR Maximization}
\label{Sec:CRPA}

The detected-channel WSR objective in~\eqref{eq:wsr} is nonconvex due to the coupling between $\mathbf F$ and $\{\mathbf u_i\}$.
All updates use the detected effective channels $\{\widehat{\mathbf H}_{k,\mathrm{eff}}\}_{k=1}^K$.

By the weighted minimum mean-square error (WMMSE) equivalence~\cite{SHI2011}, introduce per-user linear receive filters
$\mathbf U_k\in\mathbb C^{N_{r,k}\times d_k}$ and positive-definite weights
$\mathbf W_k\in\mathbb C^{d_k\times d_k}$, where $d_k$ is the number of data streams assigned to user $k$.
For clarity, the analysis considers the single-stream case $d_k=1$.
The detected-channel WSR maximization is equivalent to the following WMMSE problem:
\begin{align}
\label{eq:wmmse}
& \min_{\mathbf F,\{\mathbf U_k,\mathbf W_k\},\{\mathbf u_i\}}
\sum_{k=1}^K
\omega_k
\Big(
\mathrm{Tr}(\mathbf W_k\mathbf E_k)
-
\log\det\mathbf W_k
\Big)
\\
& \text{s.t.}\;\;
\|\mathbf F\|_F^2\le P,
\qquad
|[\mathbf u_i]_m|=1,\ \forall i,m .
\nonumber
\end{align}
The mean-square error (MSE) matrix is computed using
$\widehat{\mathbf H}_{k,\mathrm{eff}}$ as
\begin{align}
\mathbf E_k
&=
\mathbf I
-
\mathbf U_k^{H}
\widehat{\mathbf H}_{k,\mathrm{eff}}
\mathbf f_k
-
\mathbf f_k^{H}
\widehat{\mathbf H}_{k,\mathrm{eff}}^{H}
\mathbf U_k
\nonumber\\
&\quad
+
\mathbf U_k^{H}
\Big(
\widehat{\mathbf H}_{k,\mathrm{eff}}
\mathbf F\mathbf F^{H}
\widehat{\mathbf H}_{k,\mathrm{eff}}^{H}
+
\sigma^2\mathbf I
\Big)
\mathbf U_k .
\label{eq:mse}
\end{align}

A block-MM Gauss--Seidel scheme updates
$\{\mathbf U_k,\mathbf W_k\}$, $\mathbf F$, and
$\{\mathbf u_i\}$ in turn.

\noindent\textbf{(A) Receiver and Weight.}
For fixed $(\mathbf F,\{\mathbf u_i\})$, the optimal receive
filter and weight matrix are
\begin{align}
\mathbf U_k
&\leftarrow
\Big(
\widehat{\mathbf H}_{k,\mathrm{eff}}
\mathbf F\mathbf F^{H}
\widehat{\mathbf H}_{k,\mathrm{eff}}^{H}
+
\sigma^2\mathbf I
\Big)^{-1}
\widehat{\mathbf H}_{k,\mathrm{eff}}
\mathbf f_k,
\\
\mathbf W_k
&\leftarrow
\mathbf E_k^{-1}.
\end{align}

\noindent\textbf{(B) Precoder.}
For fixed
$\left(\{\mathbf U_k,\mathbf W_k\},\{\mathbf u_i\}\right)$,
define
\begin{align}
\mathbf A
&\triangleq
\sum_{k=1}^{K}
\omega_k
\widehat{\mathbf H}_{k,\mathrm{eff}}^{H}
\mathbf U_k\mathbf W_k\mathbf U_k^{H}
\widehat{\mathbf H}_{k,\mathrm{eff}},
\\
\mathbf B
&\triangleq
\Big[
\omega_1
\widehat{\mathbf H}_{1,\mathrm{eff}}^{H}
\mathbf U_1\mathbf W_1,
\ldots,
\omega_K
\widehat{\mathbf H}_{K,\mathrm{eff}}^{H}
\mathbf U_K\mathbf W_K
\Big].
\end{align}
With a Lagrange multiplier $\lambda\ge0$ for the power
constraint,
\begin{equation}
\label{eq:F-update}
\mathbf F
\leftarrow
(\mathbf A+\lambda\mathbf I)^{-1}\mathbf B,
\end{equation}
where $\lambda$ is chosen by bisection so that
$\|\mathbf F\|_F^2=P$ when the power constraint is active.

\noindent\textbf{(C) RIS Phases.}
For the update of RIS-$i$, let $f(\mathbf u_i)$ denote the
WMMSE objective in \eqref{eq:wmmse}, with
$\{\mathbf U_k,\mathbf W_k\}_{k=1}^{K}$, $\mathbf F$, and
$\{\mathbf u_j\}_{j\neq i}$ fixed, and with
$\mathbf E_k$ in \eqref{eq:mse} viewed as a function of
$\mathbf u_i$ through
$\widehat{\mathbf H}_{k,\mathrm{eff}}$.
Using Wirtinger calculus~\cite{BRANDWOOD1983}, define the
conjugate gradient as
\begin{equation}
\mathbf g_i^{(t)}
\triangleq
2
\left.
\frac{\partial f(\mathbf u_i)}
{\partial\mathbf u_i^{\ast}}
\right|_{\mathbf u_i=\mathbf u_i^{(t)}} .
\label{eq:conjugate-gradient}
\end{equation}
Assuming that the block gradient is Lipschitz continuous with
constant $L_i>0$, a quadratic majorizer around
$\mathbf u_i^{(t)}$ is given by
$f(\mathbf u_i^{(t)})
+\mathcal M_i(\mathbf u_i;\mathbf u_i^{(t)})$, where
\begin{align}
\mathcal M_i
(\mathbf u_i;\mathbf u_i^{(t)})
&\triangleq
\operatorname{Re}
\left\{
{\mathbf g_i^{(t)}}^{H}
(\mathbf u_i-\mathbf u_i^{(t)})
\right\}
\nonumber\\
&\quad
+
\frac{L_i}{2}
\|\mathbf u_i-\mathbf u_i^{(t)}\|_2^2 .
\label{eq:majorizer}
\end{align}
Since $\|\mathbf u_i\|_2^2=N_i$ over the unit-modulus
constraint set, minimizing this majorizer reduces to the
phase-alignment problem
\begin{equation}
\label{eq:phase-align}
\widetilde{\mathbf u}_i
\in
\arg\max_{|[\mathbf u_i]_m|=1}
\operatorname{Re}
\left\{
{\mathbf c_i^{(t)}}^{H}\mathbf u_i
\right\},
\qquad
\mathbf c_i^{(t)}
\triangleq
L_i\mathbf u_i^{(t)}
-
\mathbf g_i^{(t)} .
\end{equation}
The tentative solution is obtained elementwise in closed form:
\begin{equation}
\label{eq:exp-arg}
[\widetilde{\mathbf u}_i]_m
=
\exp
\left(
j\arg
\big(
[\mathbf c_i^{(t)}]_m
\big)
\right),
\:\:
\forall m.
\end{equation}
The backtracking parameter $L_i$ is selected so that the
majorization condition
\begin{align}
f(\widetilde{\mathbf u}_i)
\le
f(\mathbf u_i^{(t)})
+
\mathcal M_i
(\widetilde{\mathbf u}_i;\mathbf u_i^{(t)})
\label{eq:mm-condition}
\end{align}
holds at the tentative update.
Starting from $L_i^{(0)}>0$, $L_i$ is multiplied by
$\eta>1$, and $\widetilde{\mathbf u}_i$ is recomputed until
\eqref{eq:mm-condition} is satisfied, after which
$\mathbf u_i^{(t+1)}
\leftarrow
\widetilde{\mathbf u}_i$.
Thus, the unit-modulus constraint is satisfied by construction
without an iterative manifold retraction.
For continuous RIS phases, the block-MM update monotonically
decreases the WMMSE objective and, equivalently, monotonically
improves the detected-channel WSR.

\textbf{Discrete phase shifts.}
For practical RIS hardware, the formulation also considers
$b$-bit phase quantization:
\begin{equation}
\Theta_b
\triangleq
\left\{
0,
\frac{2\pi}{2^b},
\dots,
\frac{2\pi(2^b-1)}{2^b}
\right\},
\qquad
[\mathbf u_i]_m
\in
\{e^{j\theta}:\theta\in\Theta_b\}.
\end{equation}
The quantized CRPA candidate is obtained by nearest-neighbor
phase quantization:
\begin{align}
[\widetilde{\mathbf u}_{i,b}]_m
&=
\exp
\left(
j\,
\mathsf Q_b
\big(
\arg([\mathbf c_i^{(t)}]_m)
\big)
\right),
\label{eq:quant-update}
\\
\mathsf Q_b(\phi)
&\triangleq
\arg\min_{\theta\in\Theta_b}
\left|
e^{j\phi}-e^{j\theta}
\right|.
\end{align}

\begin{figure}[t]
    \centering
    \includegraphics[width=0.95\linewidth]{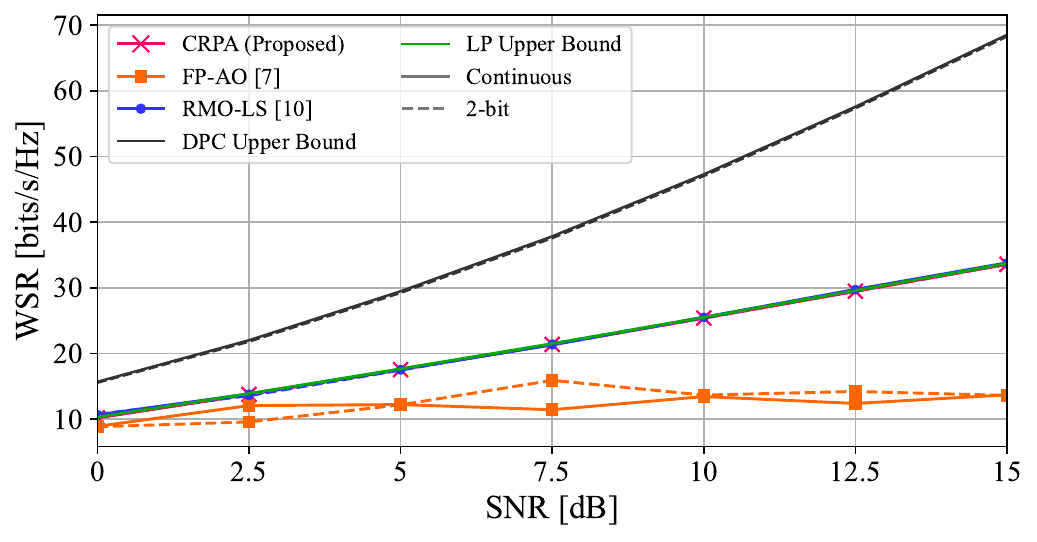}
    \vspace{-9pt}
    \caption{WSR versus SNR with $K{=}5$, $M{=}10$, and $p_{\mathrm{blk}}{=}0.1$ under continuous and 2-bit quantized RIS phases.}
    \vspace{-10pt}
    \label{fig:SNR_WSR}
\end{figure}

\noindent Unlike the continuous-phase update, quantization may break
the exact block-MM monotonicity.
Therefore, the quantized candidate is accepted only if
\begin{equation}
f(\widetilde{\mathbf u}_{i,b})
\le
f(\mathbf u_i^{(t)}).
\label{eq:quant-acceptance}
\end{equation}
Otherwise, $L_i$ is increased and the candidate is recomputed.
This yields an objective-safeguarded quantized CRPA update.

\textbf{Gradient computation.}
Let
$\mathbf Q=\mathbf F\mathbf F^{H}$,
$\boldsymbol\Psi_k
=\mathbf U_k\mathbf W_k\mathbf U_k^{H}$,
and
$\mathbf C_k
=\mathbf U_k\mathbf W_k\mathbf f_k^{H}$.
For RIS-$i$, define the set of UEs whose detected available RIS
set includes RIS-$i$ as
\begin{equation}
\widehat{\mathcal K}_i
\triangleq
\{k:i\in\widehat{\mathcal A}_k\}.
\end{equation}
Using matrix calculus and the linearity of
$\widehat{\mathbf H}_{k,\mathrm{eff}}$ with respect to
$\mathbf u_i$, the gradient is
\begin{align}
\mathbf g_i
&=
2
\sum_{k\in\widehat{\mathcal K}_i}
\omega_k
\operatorname{diag}
\left(
\mathbf H_{r,i\to k}^{H}
\left(
\boldsymbol\Psi_k
\widehat{\mathbf H}_{k,\mathrm{eff}}
\mathbf Q
-
\mathbf C_k
\right)
\mathbf G_i^{H}
\right).
\label{eq:gradient}
\end{align}

Unlike manifold-gradient methods that require retraction and line search, CRPA directly aligns each RIS phase vector with a closed-form block-MM direction and uses only a scalar backtracking parameter for surrogate validity.

\section{Experimental Results}
\label{sec:exp}

\subsection{Simulation Setup}

\textbf{Carrier/arrays:}
The simulations use a 28-GHz downlink mmWave MU--MIMO system with an $N_t{=}16$-element BS uniform linear array (ULA), $K$ UEs each equipped with $N_{r,k}{=}4$ antennas, and $M$ RISs each comprising $N_i{=}16$ reflecting elements.
All arrays use $\lambda/2$ spacing.
Unless otherwise stated, WSR is evaluated for both continuous RIS phases and 2-bit quantized phases.

\textbf{Geometry and channel/blockage:}
The BS is located at the origin.
In each Monte Carlo drop, RISs and UEs are placed at random azimuths, with BS--RIS and BS--UE distances uniformly drawn from 15--40 m and 10--50 m, respectively.
UE locations are redrawn until at least one RIS is within 15 m of the UE.
On each subcarrier, each BS--UE, BS--RIS, and RIS--UE link follows a Rician model with $K_R=5$, geometry-based line-of-sight (LoS) steering, aggregate non-line-of-sight (NLoS) multipath represented by i.i.d. Rayleigh fading, and log-distance pathloss $L(d)=61.4+20\log_{10}(d)$ dB for $d$ in meters, corresponding to the 28-GHz close-in LoS model~\cite{deng201528GHz}.
The SNR denotes the received SNR after pathloss with noise variance $\sigma^2$.
Each UE--RIS cascaded path is independently blocked with probability $p_{\mathrm{blk}}$, defining the true available RIS set $\mathcal A_k$.

\textbf{Sensing and Monte Carlo:}
Blockage detection uses indexed ZC pilots with $(\ell,q)=(63,25)$, CP length $L_{\mathrm{cp}}=8$, and the all-pass RIS sensing state $\mathbf u_i^{(0)}=\mathbf 1_{N_i}$.
Detection thresholds follow the interference-aware NP rule in Section~\ref{Sec:detect}.
Each point is averaged over $N_{\mathrm{MC}}=100$ realizations.

\begin{figure}[t]
    \centering
    \includegraphics[width=0.95\linewidth]{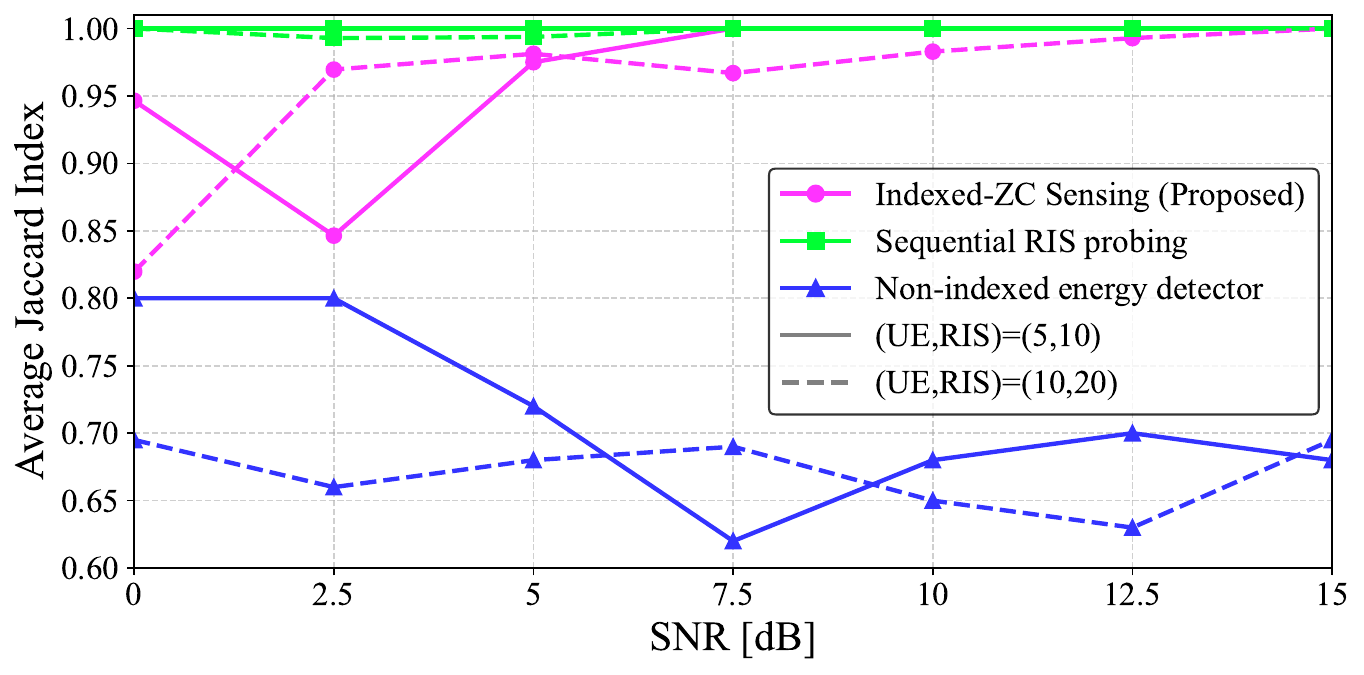}
    \caption{Jaccard index for blockage detection versus SNR under $\alpha=10^{-3}$ with indexed Zadoff--Chu pilots ($\ell$=63, $L_{\mathrm{cp}}$=8).}
    \vspace{-10pt}
    \label{fig:BDR_SNR}
\end{figure}

\subsection{Performance of Per-RIS Indexed Blockage Detection}

Fig.~\ref{fig:BDR_SNR} shows the average Jaccard index $J_k$ for $(K,M)=(5,10)$ and $(10,20)$.
The proposed indexed-ZC detector performs close to sequential RIS probing, which serves as a high-overhead reference by probing RISs one at a time, while substantially outperforming the non-indexed energy detector.
This demonstrates the advantage of per-RIS pilot indexing under Rician fading and inter-RIS interference.

\subsection{WSR Performance under Continuous and Quantized RIS Phases}

Given the detected available RIS sets $\{\widehat{\mathcal A}_k\}_{k=1}^{K}$, the evaluation considers both continuous and 2-bit quantized RIS phase shifts.
For the continuous-phase case, each RIS element can take an arbitrary phase in $[0,2\pi)$.
For the 2-bit quantized case, the feasible phase set is
$\Theta_2=\{0,\frac{\pi}{2},\pi,\frac{3\pi}{2}\}$.
CRPA directly applies the closed-form phase-alignment update in the continuous case, while in the 2-bit case it applies the nearest-neighbor quantizer \eqref{eq:quant-update} after each phase-alignment step.
All users receive equal weights, $\omega_k=1$.

Baselines are \textbf{CRPA (Proposed)}, \textbf{FP-AO}~\cite{DAMPAHALAGE2021}, \textbf{RMO-LS}~\cite{NIU2022}, and the \textbf{LP/DPC upper bounds}.
For the continuous-phase comparison, each algorithm is evaluated using its continuous RIS phase solution.
For the 2-bit comparison, RIS phases produced by each method are quantized elementwise to $\Theta_2$ before objective evaluation.
The LP/DPC upper bounds represent relaxed linear-precoding and dirty-paper-coding upper bounds computed with the same effective channels.
All methods run up to 200 iterations.

% Fig.~\ref{fig:SNR_WSR} compares WSR performance under continuous and 2-bit quantized RIS phases.
% CRPA closely tracks the LP upper bound across the SNR range, while the DPC bound remains an ideal non-linear precoding benchmark.
% FP-AO degrades under multi-RIS coupling and discrete phase constraints, whereas RMO-LS remains competitive but requires slower iterative search.
% Fig.~\ref{fig:SNR_ERR} further shows that WSR improves with blockage-detection accuracy, confirming that accurate RIS-path availability sensing is critical; under conservative false-positive suppression, residual errors mainly act as false negatives.

Fig.~\ref{fig:SNR_WSR} compares WSR performance under continuous and 2-bit quantized RIS phases.
CRPA closely tracks the LP upper bound across the SNR range, while the DPC bound serves as an ideal nonlinear-precoding benchmark.
The 2-bit result incurs only moderate loss relative to continuous phase control, indicating robustness to finite-resolution RIS hardware.
FP-AO shows weaker performance in this multi-RIS setting, whereas RMO-LS remains competitive but converges more slowly.

\subsection{Convergence and Complexity}
\begin{figure}[t]
    \centering
    \subfloat[$\mathrm{SNR}=5$ dB]{%
        \includegraphics[width=0.98\linewidth, trim=0 10 0 0, clip]{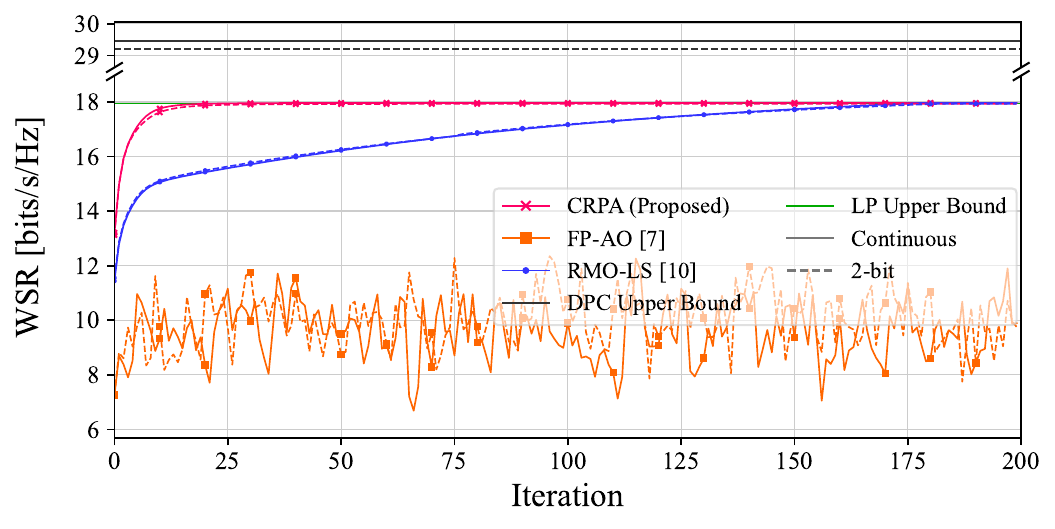}
        \vspace{-12pt}
    }\\ \vspace{-6pt}
    \subfloat[$\mathrm{SNR}=15$ dB]{%
        \includegraphics[width=0.98\linewidth, trim=0 10 0 0, clip]{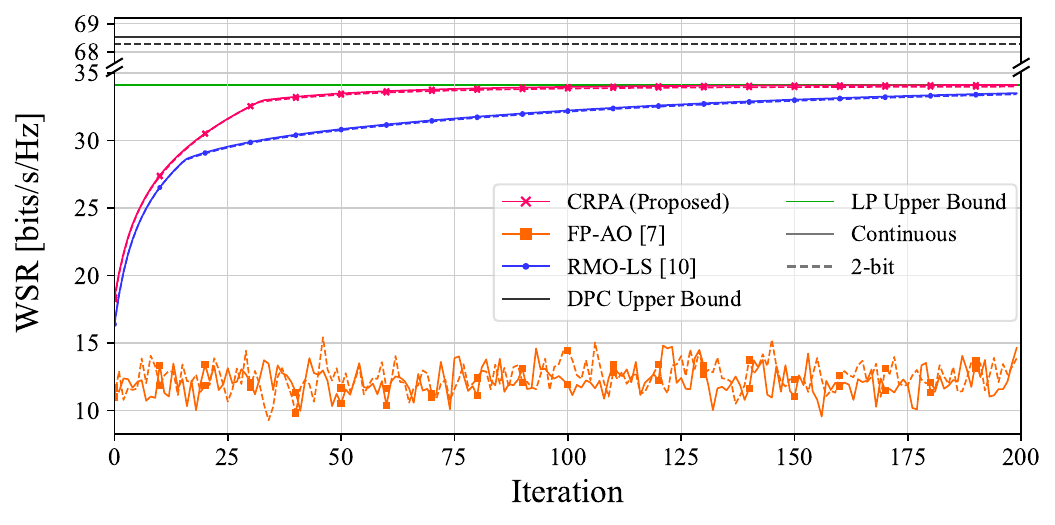}
        \vspace{-12pt}
    }
    \caption{WSR versus iteration count for $K{=}5$, $M{=}10$, and $p_{\mathrm{blk}}{=}0.1$ under continuous and 2-bit quantized RIS phases.}
    \vspace{-12pt}
    \label{fig:ITR_WSR}
\end{figure}

Fig.~\ref{fig:ITR_WSR} shows the WSR convergence at $\mathrm{SNR}=5$ and $15$ dB.
CRPA converges rapidly for both continuous and 2-bit RIS phases; in the quantized case, backtracking prevents WSR degradation caused by nearest-neighbor phase projection.
On average, CRPA reaches 95\% of its converged WSR within about 12 iterations, while RMO-LS requires substantially more iterations due to repeated line searches.
The RIS-update cost of CRPA is $O(\sum_i N_i)$, with only elementwise quantization added in the 2-bit case, whereas RMO-LS costs $O((1+T_{\mathrm{ls}})\sum_i N_i)$ with an average line-search count $T_{\mathrm{ls}}\approx35$--$45$.
Thus, CRPA provides faster convergence and lower RIS-update complexity, especially under practical low-resolution RIS control.

\section{Conclusion}
This paper presents a sensing-driven blockage-aware multi-RIS framework for mmWave smart radio environments.
Indexed ZC synchronization allows each UE to identify its available RIS set under Rician fading and inter-RIS interference, enabling joint BS precoder and RIS phase optimization.
CRPA provides closed-form unit-modulus updates for continuous phases and objective-safeguarded quantized updates for finite-resolution RISs.
Simulations confirm accurate RIS-path sensing, higher WSR, and faster convergence than multi-RIS baselines.
Future work will address dynamic environments with time-varying blockage, channel aging, and sensing-to-control latency.

\bibliographystyle{IEEEtran}
\bibliography{ref}

\end{document}